\documentclass[fleqn,usenatbib]{mnras} 

\usepackage{newtxtext,newtxmath}

\usepackage[T1]{fontenc}
\usepackage{ae,aecompl}

\usepackage{graphicx}	
\usepackage{amsmath}	
\usepackage{grffile}    
\usepackage{breqn}

\def\cm{{\rm\thinspace cm}}
\def\km{{\rm\thinspace km}}
\def\s{{\rm\thinspace s}}

\def\g{{\rm\thinspace g}}
\def\erg{{\rm\thinspace erg}}
\def\kmps{\hbox{${\rm\km\s^{-1}\,}$}}
\def\ergps{\hbox{${\rm\erg\s^{-1}\,}$}}
\def\gpcm3{\hbox{${\rm\g\cm^{-3}\,}$}}

\def\Msol{\hbox{${\rm\thinspace M_{\odot}}$}}

\def\hii{\hbox{H{\sc ii}\,}}

\title[\hii region expansion] 
{How D-type \hii region expansion depends on numerical resolution}

\author[J.~M.~Pittard et al. ]{J.~M.~Pittard\thanks{E-mail:
    j.m.pittard@leeds.ac.uk}, M.~M.~Kupilas and C.~J.~Wareing\\
School of Physics and Astronomy, University of
       Leeds, Woodhouse Lane, Leeds LS2 9JT, UK\\   
}

\date{Accepted 2021 December 16. Received 2021 December 16; in original form 2021 November 25}

\pubyear{2022}

\begin{document}
\label{firstpage}
\pagerange{\pageref{firstpage}--\pageref{lastpage}}
\maketitle

\begin{abstract}
  We investigate the resolution dependence of \hii regions expanding
  past their Str\"{o}mgren spheres. We find that their structure and
  size, and the radial momentum that they attain at a given time, is
  in good agreement with analytical expectations if the Str\"{o}mgren
  radius is resolved with $dr \leq 0.3\,R_{\rm st}$. If this is not
  satisfied, the radial momentum may be over- or under-estimated by
  factors up to 10 or more. Our work has significance for the amount of
  radial momentum that a \hii region can impart to the ambient medium
  in numerical simulations, and thus on the relative importance of
  ionizing feedback from massive stars.
\end{abstract}




\section{Introduction}
\label{sec:intro}
Massive stars affect their surroundings through their ionizing
radiation, powerful winds, and supernova (SN) deaths. These inputs
heat and accelerate nearby gas, and can both compress and disperse gas
clouds \citep[e.g.][]{Rogers:2013,Kim:2018,Wareing:2018}. As such, massive stars
are recognized as key agents influencing star formation on both local
and galactic scales. In recent years, attention has focused on the
radial momentum that \hii regions, wind-blown bubbles, and supernova
remnants can inject into the interstellar medium, since this
determines the amplitude of gas motions which limit gravitational
condensation and collapse \citep[e.g.][]{Shetty:2012}.

Initial implementations of supernova feedback in galaxy and cosmology
simulations used an energy injection approach, and suffered from an
``over-cooling'' problem caused by insufficient numerical resolution
\citep[e.g.][]{Katz:1992}. Only in the latest prescriptions has
SN-driven feedback become independent of numerical resolution
\citep[e.g. the FIRE-2 algorithm implemented
by][]{Hopkins:2018}. Similarly, wind feedback has not always been
adequately resolved. \citet*{Pittard:2021} showed that the wind
injection radius must be below some maximum value, $r_{\rm inj,max}$, in order for the
bubble momentum to closely agree with analytical
predictions. Agreement within 25 per cent was obtained when $r_{\rm
  inj} = 0.1\,r_{\rm inj,max}$, and within 10 per cent when $r_{\rm
  inj} \lesssim 0.02\,r_{\rm inj,max}$.

We now turn our attention to ionizing feedback, which creates \hii
regions around massive stars. Much numerical modelling of \hii regions
exists in the literature, but we find that not all work has the
necessary numerical resolution to capture the correct growth of the
\hii region and its radial momentum. In this work we examine how the
development of the \hii region depends on numerical resolution. We
focus only on the ionizing feedback, so that other effects, such as
the impact of the wind, for example, do not complicate the matter.  In
Sec.~\ref{sec:hii_regions} we discuss the essential theory of \hii
regions. In Sec.~\ref{sec:calcs} we describe our calculations and our
implentation of the photoionization microphysics.  In
Sec.~\ref{sec:results} we present our results. We summarize and
conclude in Sec.~\ref{sec:summary}.

\section{\hii region essential features}
\label{sec:hii_regions}
For simplicity we consider a star that emits ionizing photons at a
constant rate $\dot{S}$ into a neutral medium of constant density and
pressure. We assume that there are no dust grains or magnetic
field. In reality, radiation pressure on dust and the dynamics of
dust-gas coupling can be important for \hii region dynamics
\citep{Akimkin:2017}.

The ionizing photons ionize the neutral gas, and drive an
ionization front that moves at a velocity $v_{\rm IF}$. Throughout
this work we assume that any recombination to the ground state of
hydrogen creates an ionizing photon with a very short mean free path
and thus creates an ionization at roughly the same location
\citep{Osterbrock:1989}. In this ``on-the-spot'' approximation the
case B recombination coefficient is appropriate. If the neutral gas is
molecular a dissociation front also moves outwards at a velocity
$v_{\rm DF}$. At early times these fronts are coincident
\citep{Krumholz:2007}.

The ionization front expands very rapidly at first, and is known as
R-type \citep{Kahn:1954}. Its radius increases as
\begin{equation}
r_{\rm IF} = R_{\rm st}\left(1 - e^{-t/t_{\rm rec}}\right)^{1/3}, 
\end{equation} 
where $R_{\rm st}$ is the Str\"{o}mgren radius given by \citep{Stromgren:1939}
\begin{equation}
R_{\rm st} = \left(\frac{3 \dot{S}}{4 \upi \alpha_{\rm rr}^{\rm B}n_{\rm
      H}^{2}}\right)^{1/3}.
\end{equation}  
\noindent
The recombination timescale,
$t_{\rm rec} = 1/\alpha_{\rm rr}^{\rm B} n_{\rm H}$, is roughly the
timescale for this first phase.  The case B recombination coefficient
is $\alpha_{\rm rr}^{\rm B} \approx 2.59\times10^{-13}(T/10^{4}\,{\rm
  K})^{-0.7}\,{\rm cm^{3}\,s^{-1}}$
\citep{Osterbrock:1989,Rijkhorst:2006}, and $n_{\rm H}$ is the total
hydrogen nucleon number density (molecular plus atomic plus ionized).

The ionized gas has a substantially higher pressure than the
surrounding neutral gas (mostly due to the increase in temperature,
but also because of the increase in number density). This pressure
increase causes the ionized gas to expand after a sound-crossing
timescale. Around this time the ionization front changes from R-type to
D-type. Because the ionization front moves at subsonic speed relative
to the ionized gas but at supersonic speed relative to the neutral
gas, it drives a shock front into the surrounding
medium and sweeps up a dense shell of neutral material.

For $\dot{S}\gtrsim 10^{47}\,{\rm s^{-1}}$, the molecular
hydrogen dissociation front does not have a significant effect on the
dynamics, as it remains trapped between the ionization front and the
shock front \citep{Hosokawa:2005,Krumholz:2007}. During the D-type
expansion, \citet{Ritzerveld:2005} found that direct photons still
dominate over diffuse ones and the on-the-spot approximation remains
valid.

The shock radius in this second phase evolves as \citep{Spitzer:1978,Hosokawa:2006,Bisbas:2015}
\begin{equation}
\label{eq:rrad}  
R_{\rm sh} = R_{\rm st}\left(1 +
  \frac{7}{4}\sqrt{\frac{4}{3}}\frac{c_{\rm i}t}{R_{\rm st}}\right)^{4/7},  
\end{equation}  
\noindent where $c_{\rm i}$ is the isothermal sound speed of the ionized gas. The
shock velocity is given by
\begin{equation}
v_{\rm sh} = c_{\rm i} \sqrt{\frac{4}{3}\left(\frac{R_{\rm st}}{R_{\rm
        sh}}\right)^{3/2} - \frac{\mu_{\rm i}T_{\rm 0}}{2\mu_{\rm
      0}T_{\rm i}}},
\end{equation}  
where $T$ is the gas temperature, $\mu$ is the mean molecular weight,
and subscripts ``0'' and ``i'' indicate the ambient and ionized medium
respectively. The radial momentum of the shell swept-up by the expanding \hii
region is
\begin{equation}
\label{eq:mtm}  
p_{\rm sh} = \frac{4\upi}{3}(R_{\rm sh}^{3} - R_{\rm
  st}^{3})\rho_{0}v_{\rm s}.
\end{equation}  
Eventually, the \hii region attains pressure equilibrium with its
surroundings. The radius at this time is \citep*{Raga:2012}
\begin{equation}
\label{eq:rstag}
R_{\rm stag} = R_{\rm st}
\left(\frac{8}{3}\right)^{2/3}\left(\frac{c_{\rm i}}{c_{\rm 0}}\right)^{4/3},
\end{equation}  
where $c_{0}$ is the isothermal sound speed in the neutral medium.

\section{The calculations}
\label{sec:calcs}
The Euler equations of gas dynamics for a spherically symmetric, inviscid
and non-heat-conducting fluid may be written in Lagrangian coordinates in
conservative form as follows (for the conservation of mass, momentum and
energy, respectively):
\begin{eqnarray}
  \label{eq:hydro_mass}
  \frac{\partial}{\partial t}\left(\frac{1}{\rho}\right) - \frac{\partial (r^{2}u)}{\partial
  m} = 0, \\ 
  \label{eq:hydro_mtm}
  \frac{\partial u}{\partial t} + r^{2}\frac{\partial p}{\partial
  m} = 0, \\
  \label{eq:hydro_energy}
  \rho\left[\frac{\partial E_{\rm m}}{\partial t} + \frac{\partial (r^{2}up)}{\partial
  m}\right] = \dot{E}_{\rm int,v},
\end{eqnarray}  
\noindent where $\rho$ is the fluid mass density, $u$ is the velocity and
$E_{\rm m}$ is the total energy per unit mass. $m$ is the mass coordinate
defined as $dm = \rho r^{2} dr$, where $r$ is the radial
coordinate. The internal energy per unit mass $e_{\rm m} = E_{\rm m} -
u^{2}/2$, and the pressure $p = (\gamma-1)\rho e_{\rm m}$.  The source term on
the right-hand side of the energy equation, $\dot{E}_{\rm int,v}$, is
the internal energy change per unit volume, and represents cooling and
heating processes that are discussed below. 

We use a heavily modified version of
VH-1\footnote{http://wonka.physics.ncsu.edu/pub/VH-1/} to solve
Eqs.~\ref{eq:hydro_mass}-\ref{eq:hydro_energy}.  The interface values
are obtained via piecewise parabolic spatial reconstruction of the
cell-averaged quantities, with flattening as appropriate. A 2-shock
approximate Riemann solver is then used to obtain the interface
fluxes, based on averages over the domain of influence of the
characteristics. The cell-averaged quantities are then updated and a
conservative remap is used to place them back onto the original
stationary Eulerian grid. The method is third-order accurate in space
for smooth parts of the flow, and first-order at shocks. A Courant
number of 0.6 is used.

An advected scalar is used to track the hydrogen ionization fraction,
$y$. Advected scalars are unchanged by the Lagrangian step but are
modified during the remap step. The neutral fraction $x = 1 - y$. The
total H number density $n_{\rm H} = \rho/\mu_{\rm H}$, where
$\mu_{\rm H}$ is the mean mass per H nucleon. The number density of
neutral hydrogen nucleons is $n_{\rm HI} = x n_{\rm H}$, and the number density
of ionized hydrogen is $n_{\rm HII} = y n_{\rm H}$. To calculate the
electron number density, $n_{\rm e}$, we assume that He is singly
ionized whenever H is \citep{Mackey:2015}, and that C is always singly
ionized due to the interstellar UV field \citep{Rijkhorst:2006}. We
assume that all of the metals are Carbon. The electron number density
is then given by $n_{\rm e} = y(n_{\rm H} + n_{\rm He}) + n_{\rm C}$,
where $n_{\rm He}$ and $n_{\rm C}$ are the Helium and Carbon number
densities, respectively. We assume mass fractions $X_{\rm H}=0.7381$, $X_{\rm He}=0.2485$,
and $X_{\rm C}=0.0134$ for the abundances \citep[cf.][]{Grevesse:2010}.

Changes to the ionization of the gas
and heating/cooling processes are included via an operator split
step. The rate of change of the ionization fraction and the internal
energy per unit volume are: 
\begin{eqnarray}
\label{eq:ydot}  
  \dot{y} = A_{\rm pi}(1-y) + A_{\rm ci}n_{\rm H}y(1-y) - \alpha_{\rm
  rr}^{\rm B}n_{\rm H}y^{2},\\
\label{eq:eintdot}
  \dot{E}_{\rm int,v} = (\rho/m_{\rm H}) \Gamma + \mathcal{G}_{\rm ph} - (\rho/m_{\rm H})^{2}\Lambda(T) - n_{\rm e}n_{\rm
  HII}\Lambda_{\rm rec}.
\end{eqnarray} 
In Eq.~\ref{eq:ydot}, the terms on the right hand side are due to
photoionization, collisional ionization and recombination. In
Eq.~\ref{eq:eintdot}, the terms on the right hand side are due to
background heating, heating due to the photoionization process, gas
cooling and recombination cooling.

The ionizing radiation model uses a photon conservative scheme. The
photoionization rate coefficient, $A_{\rm pi}$, depends on the rate of
ionizing photons entering the cell minus the rate leaving. The
photoionization rate within the cell is given by
\begin{equation}
\dot{N}_{\rm ion}= \dot{S} e^{-\tau}(1.0 - e^{-d\tau}), 
\end{equation}
\noindent where $\tau$ is the optical depth to ionizing photons from the star to the
inner edge of the cell, and $d\tau$ is the optical depth to ionizing
photons in the cell. The optical depth
\begin{equation}
\tau = \int (1-y)\,\sigma\,n_{\rm H} dl,  
\end{equation}  
where $\sigma = 6.3\times10^{-18}\,{\rm cm^{2}}$ is the
photoionization cross-section for neutral H at the ionizing threshold
and $dl$ is the path length. We then have $A_{\rm pi} =
\dot{N}_{\rm ion}/n_{\rm HI}V$, 
where $V$ is the cell volume. The
collisional ionization rate coefficient is given by $A_{\rm ci} =
5.84\times10^{-11} \sqrt{y} \exp(-13.6/kT)\,{\rm cm^{3}\,s^{-1}}$ for gas at
temperature $T$. 

We assume that each absorption of an ionizing photon results in a
photoelectron with an energy $e_{\Gamma} = 2.4$\,eV
\citep{Whalen:2006}. These heat the gas, giving a heating rate per
unit volume $\mathcal{G}_{\rm ph} = e_{\Gamma}\dot{N}_{\rm ion}/V$.
For the recombination cooling we use
$\Lambda_{\rm rec} = 6.1\times10^{-10}kT^{0.11} {\rm
  \,erg\,cm^{3}\,s^{-1}}$ if $T \gtrsim 100$\,K
\citep{Osterbrock:1989}. The cooling curve, $\Lambda(T)$, is
constructed from 3 separate parts \citep[see
also][]{Wareing:2017a,Wareing:2017b,Kupilas:2021}.  At low
temperatures ($T < 10^{4}$\,K) we use a fit to the data in
\citet{Koyama:2002}, corrected by \citet{Vazquez-Semadeni:2007}:
\begin{equation}
\frac{\Lambda}{\Gamma}=10^{7}\,\exp\left(\frac{-1.184\times 10^{5}}{T
    + 1000}\right) + 0.014 \sqrt{T} \exp\left(\frac{-92}{T}\right).
\end{equation}
For $10^{4}\leq T/{\rm K} <
10^{7.6}$, $\Lambda(T)$ is constructed using data from {\sc
  CLOUDY} v10.0 \citep{Gnat:2012}. For $T \geq 10^{7.6}$\,K, we use
data from the {\sc MEKAL} plasma
emission code \citep*{Mewe:1995}. We use a constant heating coefficient
($\Gamma = 2 \times 10^{-26} \,{\rm erg\,s^{-1}}$).

A temperature-dependent average particle mass, $\mu$, is used. In the
molecular phase $\mu = 2.36$, while $\mu = 0.61$ in ionized gas. The
value of $\mu$ is determined from a look-up table of values of
$p/\rho$ \citep{Sutherland:2010}. The ratio of specific
heats is set as $\gamma = 5/3$ at all temperatures.

In the operator split step we integrate $x$ and $E_{\rm int,v}$ using
the {\sc CVODE} solver from the {\sc sundials} v5.8.0 numerical
library\footnote{https://computing.llnl.gov/projects/sundials}. {\sc
  CVODE} is a sophisticated solver that automatically detects
stiffness. Like \citet{Mackey:2012}, we find that the numerical
integration is more stable if $x$ rather than $y$ is
integrated. Because the ray-tracing is performed once per step, the
photon conservation is first-order accurate in time, and our
photoionization algorithm is the same as method A2 in
\citet{Mackey:2012}. We also set the same error tolerances for the
{\sc CVODE} solver (a relative error of $10^{-4}$ and absolute errors
of $10^{-12}$ and $10^{-17}$ for $x$ and $E_{\rm int,v}$,
respectively). Following \citet{Mackey:2012}, we also limit the
timestep of the microphysics to
\begin{equation}
\Delta t = {\rm min}\,\left( K_{1} t_{\rm rec}, K_{2}\frac{E_{\rm
      int,v}}{|\dot{E}_{\rm int,v}|}, K_{3}\frac{{\rm
      max}(0.05,1-y)}{|\dot{y}|}, K_{4}\frac{1}{|\dot{y}|}\right).
\end{equation}  
In all of our calculations we set
$K_{1}=K_{4}=\infty$ and $K_{2}=K_{3}=0.3/\tau_{\rm cell}$, where
$\tau_{\rm cell}$ is the initial optical depth of each grid cell.
We use the smallest of the Courant-limited and microphysics-limited
timesteps to advance both the hydrodynamics and the microphysics
(i.e. we do not super-sample the microphysics).

Other ionization schemes are available that are more sophisticated
than our scheme. These include the second-order explicit method A3 in
\citet{Mackey:2012}, and implicit schemes, such as ${\rm C^{2}}$-ray
\citep{Mellema:2006} and method A1 in \citet{Mackey:2012}.
However, because of the nature of the PPMLR hydrodynamics scheme used
in this work, a first-order photoionization scheme is appropriate
here. We do not expect our conclusions to be affected by our choice of
scheme.

Naively, one might expect that the Str\"{o}mgren radius should be
resolved in order that the \hii region expand correctly.  Therefore,
our focus is around this numerical resolution and we define
\begin{equation}
\label{eq:chi}  
\chi = \frac{dr}{R_{\rm st}},  
\end{equation}  
where $dr$ is the width of the grid cells. We then vary the value of $\chi$
in our simulations.

\section{Results}
\label{sec:results}
We adopt the following set of parameters for all of our simulations.
We assume that $\dot{S} = 10^{49} \,{\rm s^{-1}}$ and
$\rho_{0} = 2\times10^{-21}\,{\rm g\,cm^{-3}}$
($n_{\rm H} = \rho_{0}/\mu_{\rm H} = 884\,{\rm cm^{-3}}$, given
a mean mass per H nucleon $\mu_{\rm H}=2.26\times10^{-24}$\,g). Our
adopted value of $\rho_{0}$ determines that
$\mu_{0} = 2.36$ and $T_{0} = 21$\,K. The pressure of
the ambient gas, $p_{0} = 1.48\times10^{-12}\,{\rm dyn \,cm^{-2}}$ (or
$p_{0}/k = 1.07\times10^{4}\,{\rm K \,cm^{-3}}$). This then gives
$c_{0} = 5.3\times10^{4}\,{\rm cm\,s^{-1}}$. We find that the
temperature of the ionized gas $T_{\rm i} \approx 8100$\,K, giving
$c_{\rm i} \approx 1.04\times10^{6}\,{\rm cm\,s^{-1}}$. The mean
molecular weight in the ambient and ionized gas is $\mu_{0}=2.36$ and
$\mu_{\rm i}=0.61$, respectively.

The Str\"{o}mgren radius, $R_{\rm st} = 0.702$\,pc. Due to the large
ratio of $c_{\rm i}/c_{0}$, the stagnation radius
$R_{\rm stag} \approx 70$\,pc. We evolve the simulations for $5\,$Myr,
which is a typical lifetime for a massive star with an ionizing flux
of this magnitude. Table~\ref{tab:models} lists some other details of
our models. In model {\em chi0.1}, each cell has a width
$dr=0.0702$\,pc and an optical depth to ionizing photons
$\tau_{\rm cell} = 1210$. The other models have larger cell widths and
optical depths.

\begin{table}
\begin{center}
  \caption[]{The models investigated. The columns show the model name,
    the grid resolution, the ratio of the grid resolution to the
    Str\"{o}mgren radius (Eq.~\ref{eq:chi}),
    and the measured radial momentum of the \hii region after
    5\,Myr.}
\label{tab:models}
\begin{tabular}{lccc}
\hline
  Model & $dr$ & $\chi$ & $p_{\rm sh}$ \\
        & (pc) &        & ($\Msol \,\kmps$) \\
  \hline
  {\it chi0.1} & $0.0702$ & $0.1$ & $2.78\times10^{5}$ \\
  {\it chi0.3} & $0.211$ & $0.3$ & $2.86\times10^{5}$ \\
  {\it chi1.0} & $0.702$ & $1.0$ & $3.12\times10^{5}$ \\
  {\it chi3.0} & $2.11$ & $3.0$ & $3.57\times10^{5}$ \\
  {\it chi5.0} & $3.51$ & $5.0$ & $4.03\times10^{5}$ \\
  {\it chi7.5} & $5.27$ & $7.5$ & $3.33\times10^{5}$ \\
  {\it chi10}  & $7.02$ & $10$ & $6.88\times10^{4}$ \\
  {\it chi30}  & $21.1$ & $30$ & $2.01\times10^{4}$ \\
\hline
\end{tabular}
\end{center}
\end{table}

\begin{figure*}
\includegraphics[width=8.0cm]{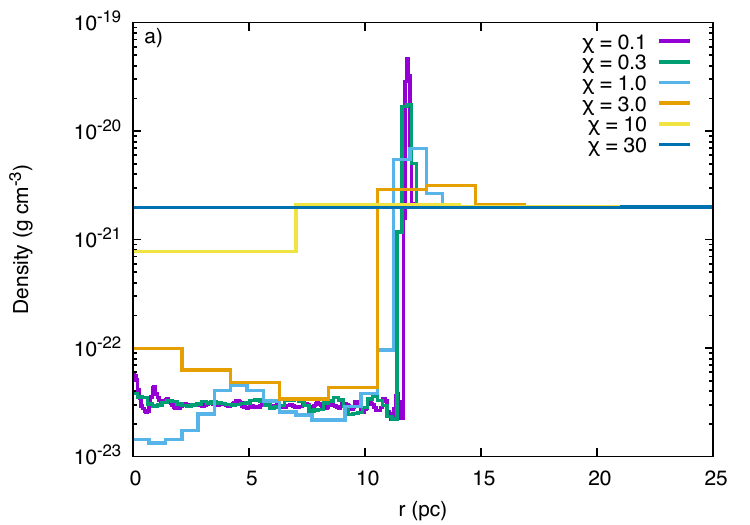}
\includegraphics[width=8.0cm]{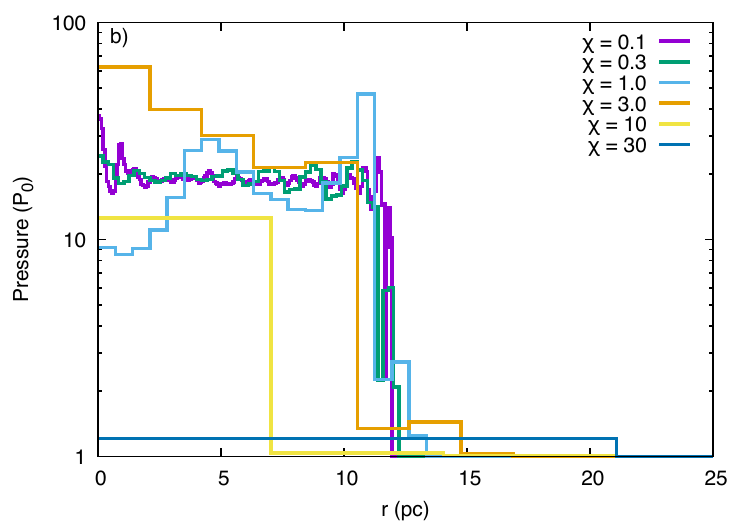}
\includegraphics[width=8.0cm]{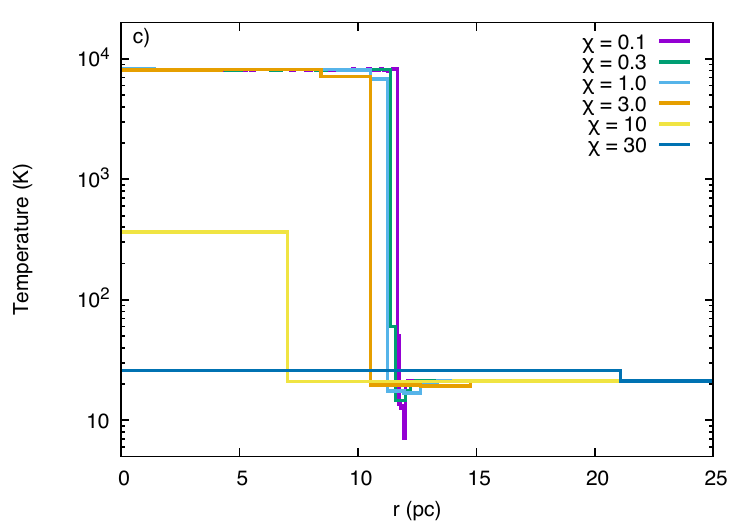}
\includegraphics[width=8.0cm]{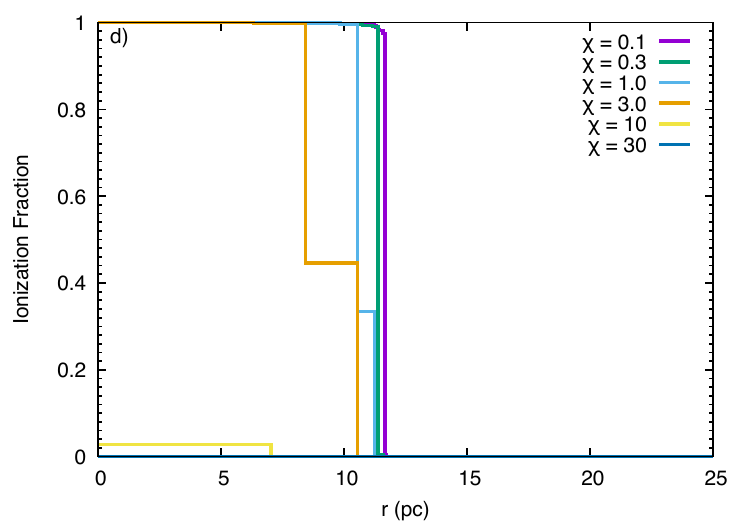}
\caption{Profiles of a) density; b) pressure; c) temperature; and d)
  ionization fraction at $t=5\,$Myr for models with $\chi = 0.1$, 0.3,
  1.0, 3.0, 10 and 30. The ambient density and temperature values are
  visible on the far right of the plots in panels a) and c). Note the
  differences in the profiles as the resolution is varied. To aid the
  reader the plot style is deliberately chosen to show steps, as this
  displays the cell averaged quantity over the radii encompassed by
  each cell. In model {\em chi30} the cell width is 21.1\,pc, so
  values from only two cells are visible.}
\label{fig:hii_profiles}
\end{figure*}

\subsection{\hii region profiles}
Fig.~\ref{fig:hii_profiles} shows profiles of density, pressure,
temperature and ionization fraction at $t=5$\,Myr for each of our
models. In model {\em chi0.1}, we can clearly see the dense shell (at
$r\approx12$\,pc) swept-up by the expanding \hii region. Although the
maximum density in the shell is not converged, with models with
smaller values of $\chi$ showing higher values, the global properties
are converged. The \hii region is still a factor of 20 over-pressured
with respect to the ambient medium at this time. Due to the
compression of the gas in the swept-up shell, the temperature drops
below 10\,K. The ionization fraction of the gas drops away from unity
only near the edge of the \hii region. Waves within the \hii region
are also visible. These cause the density, velocity and pressure to
oscillate, but the temperature and ionization fraction are largely
unaffected. Waves are also seen in other work \citep[e.g. see Fig.~4
in][]{Bisbas:2015}.  In our case they may also result from the PPMLR
method employed by VH-1 where strong shocks that move slowly across
the grid are known to cause strong oscillations.

As the resolution of the models change, the profiles begin to deviate
from model {\em chi0.1}. The $\chi=1$ model matches the higher
resolution models reasonably well, and the $\chi=3$ model still
displays their main qualitative features despite not resolving the
Str\"{o}mgren radius. It is clear, therefore, that models with
$\chi>1$ may still create a \hii region. In such cases, gas in the
grid cell closest to the star becomes partially photoionized (from a
greater to a lesser degree as $\chi$ increases). This raises the cell
pressure which initiates a flow of gas out of the grid cell. The
density in the grid cell drops, which allows the ionization fraction
to increase further. In model {\em chi3.0}, this process runs-away on
a timescale determined by the decreasing sound-crossing time of the
gas as the cell changes from partially to fully ionized. The result is
that model {\em chi3.0} creates a \hii region with features
qualitatively similar to higher resolution models by $t=5$\,Myr.

In contrast, models with $\chi\gtrsim 10$ fail to create completely
ionized gas with $y=1.0$ and $T=8100$\,K at $t=5$\,Myr in the grid
cell closest to the star. The ionization fraction of this gas is 2.8\%
and 0.08\% in models {\em chi10} and {\em chi30}
respectively. Nevertheless, in both cases the partially ionized gas is
able to initiate a flow away from the star due to the pressure
difference that exists between it and the ambient gas. At $t=5$\,Myr,
the ratio of $p/p_{\rm 0}$ is 12.1 and 1.21 in models {\em chi10} and
{\em chi30}, respectively.

\begin{figure*}
\includegraphics[width=8.0cm]{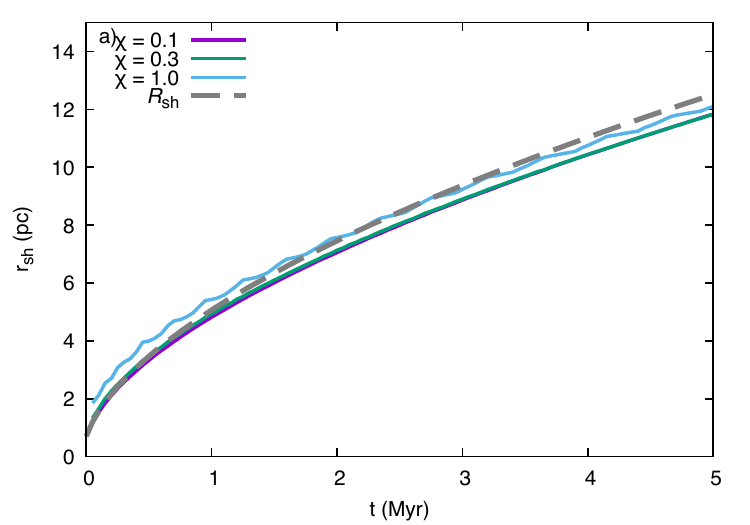}
\includegraphics[width=8.0cm]{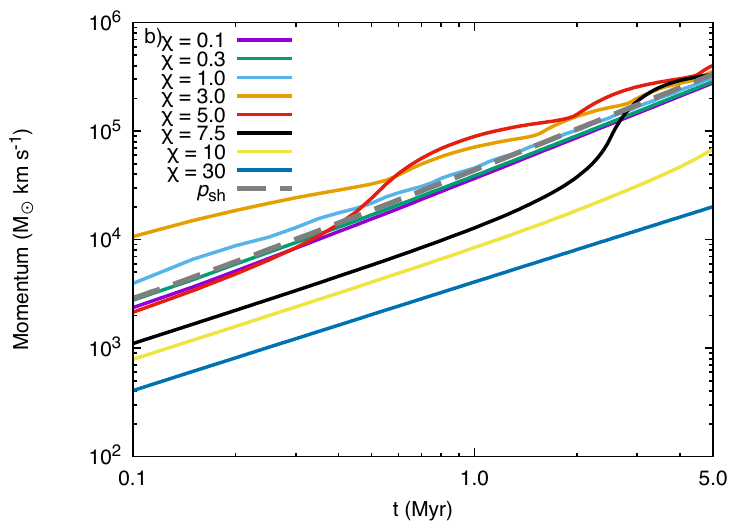}
\caption{a) The radius of the shock front as a function of time for
  models with $\chi \leq 1$. b) The radial momentum of the \hii region as a
  function of time for each model. The lines labelled $R_{\rm sh}$ and
  $p_{\rm sh}$ are calculated using Eqs.~\ref{eq:rrad}
  and~\ref{eq:mtm}, respectively.}
\label{fig:hii_evolution}
\end{figure*}

\subsection{\hii region size and momentum evolution}
Fig.~\ref{fig:hii_evolution}a) shows the shock front radius in models
with $\chi \leq 1$. The shock front position is calculated as
follows. We find the cell with the highest density and examine cells
either side to see if they have any excess mass, $\delta m$ (i.e if
$\rho > \rho_{0}$). Those cells that have excess mass are included in
the summations to obtain a mass weighted radius
($r = \sum \delta m r_{\rm cell}/\sum \delta m$, where $r_{\rm cell}$
is the radius of the centre of the grid cell). 
The shock radius from models with $\chi>1$ is very dependent on the
particular algorithm and so are not shown. For the models shown in
Fig.~\ref{fig:hii_evolution}a), the shock front radius
compares well with analytical expectations. 

Fig.~\ref{fig:hii_evolution}b) shows the radial momentum of the \hii
region, calculated by summing over {\em every} grid cell, including
those past the shock front. Careful checks were made to ensure that small random
velocity perturbations to the ambient gas due to numerical round-off
error did not make any significant contribution to the measured
momentum. Simulations with identical resolution but a different number
of grid cells also confirm that this is the case.

It is immediately clear from Fig.~\ref{fig:hii_evolution}b) that the
simulations show some complex behaviour.  Models with $\chi < 1$ are
in good agreement with analytical expectations over the whole
timescale considered. In models {\em chi1.0} and {\em chi3.0} the
radial momentum is over-estimated at early times but converges towards
the analytical solution at late times. In model {\em chi5.0} the
radial momentum is over-estimated at mid-late times. Models with
$\chi \geq 10$ show a dramatic reduction in the radial momentum at all
times considered. Interestingly, we see that the model with $\chi=7.5$
initially underestimates the radial momentum, but that there is a
rapid increase between $t=2-3$\,Myr. This timing is consistent with
the initial sound crossing time of the gas in the grid cells of
$9.7$\,Myr (an upper limit, with this timescale dropping as the gas heats). It
appears that the {\em chi10} model is also heading for a similar rapid
rise.

Table~\ref{tab:models} lists the radial momentum from each model at
$t=5$\,Myr. The radial momentum obtained from Eq.~\ref{eq:mtm} is
$p_{\rm sh} = 3.32\times10^{5}\,\Msol\,\kmps$. As already noted,
the momentum measured from model {\em chi0.1} is in good agreement with it.

\section{Summary and conclusions}
\label{sec:summary}
We have examined the effect of numerical resolution on the D-type
expansion of \hii regions. We find that a \hii region can be created,
expand, and attain a radial momentum in good agreement with analytical
predictions if the Str\"{o}mgren radius is resolved such that
$\chi = dr/R_{\rm st} \leq 0.3$. With $\chi=1.0$ the radial momentum
is overestimated at early times. Models with higher values of $\chi$
either overestimate, or significantly underestimate the radial
momentum. For $\chi=10$ and $\chi=30$, the final radial momentum
measured from our models is reduced by factors of 4 and 14,
respectively.

Not all numerical simulations in the published literature seem to
resolve the Str\"{o}mgren radius. Amongst the {\sc SILCC} group of
papers, \citet{Peters:2017} were the first to consider photoionization
feedback. The {\sc SILCC} models have a resolution of $dx =
3.9$\,pc. In their model FRWSN (which also includes wind feedback),
the sink particles are star clusters with a typical mass of
$10^{3}\,\Msol$. These clusters have an ionizing luminosity
$L_{\rm ion} \sim 10^{39}\,\ergps$, which corresponds to
$\dot{S} \sim 5\times10^{49}\,{\rm s^{-1}}$. Since the sink particles
are created above a density threshold
$\rho = 2\times10^{-20}\,{\rm g\,cm^{-3}}$, we estimate that
$R_{\rm st} \approx 0.2$\,pc. This gives $\chi \approx 20$.

In another paper, \citet{Butler:2017} describe kpc-scale zoom
simulations of a galactic disk.  The resolution is 0.5\,pc. Sink
particles are generated in cells where $n > 10^{5}\,{\rm
  cm^{-3}}$. They are born with a mass of $100\,\Msol$ and
IMF-averaged stellar evolution tracks are then followed. No accretion
takes place onto the star particles. \citet{Rosdahl:2015} shows that
$L_{\rm UV}/\Msol = 10^{36}\ergps$, so each star particle has an
ionizing flux $\dot{S} \sim 5\times10^{48}\,{\rm s^{-1}}$. The
resulting Str\"{o}mgren radius is $R_{\rm st} \sim 0.02$\,pc. This
gives $\chi\approx25$.

In both of these papers, the resolution is likely to be too low for
the \hii reigons to grow correctly (unless they are clustered
together).  We stress that these papers are simply ones that we are
familiar with; other work may suffer also from this problem. In
scanning the literature we have sometimes found it hard to determine a
value for the Str\"{o}mgren radius given the information presented.
We hope that future numerical work will explicitly demonstrate that
the Str\"{o}mgren radius is sufficiently resolved (i.e.
$\chi \lesssim 0.3$).

A further complication is that in both reality and in numerical
simulations, the \hii region is typically interacting with a very
inhomogeneous medium. In such cases the \hii region will expand more
quickly into regions of lower density, and vice-versa. While the
global behaviour of the \hii region can likely be represented by an
averaged density for the local environment, it is not immediately
clear how the radial momentum attained in such circumstances may
differ from the spherically symmetric case. Further study of such a
scenario is therefore warranted.

\section*{Acknowledgements}
We thank the referee for their helpful comments. JMP was supported by
grant ST/P00041X/1 (STFC, UK).






\begin{thebibliography}{99}
\bibitem[\protect\citeauthoryear{Akimkin et al.}{2017}]{Akimkin:2017}
Akimkin V.~V., Kirsanova M.~S., Pavlyuchenkov Ya.~N., Wiebe D.~S.,
2017, MNRAS, 469, 630  
\bibitem[\protect\citeauthoryear{Bisbas et al.}{2015}]{Bisbas:2015}
Bisbas T.~G., et al., 2015, MNRAS, 453, 1324 
\bibitem[\protect\citeauthoryear{Butler et al.}{2017}]{Butler:2017}
Butler M.~J., Tan J.~.C., Teyssier R., Rosdahl J., Van Loo S.,
Nickerson S., 2017, ApJ, 841, 82
\bibitem[\protect\citeauthoryear{Gnat \& Ferland}{2012}]{Gnat:2012}
Gnat O., Ferland G.~J., 2012, ApJS, 199, 20
\bibitem[\protect\citeauthoryear{Grevesse et al.}{2010}]{Grevesse:2010}
Grevesse N., Asplund M., Sauval A.~J., Scott P., 2010, Ap\&SS, 328, 179  
\bibitem[\protect\citeauthoryear{Hopkins et al.}{2018}]{Hopkins:2018}
Hopkins P.~F., Wetzel A., Kereš D., Faucher-Giguère C.-A., Quataert E., Boylan-Kolchin M., Murray N., Hayward C.~C., El-Badry K., 2018, MNRAS, 477, 1578  
\bibitem[\protect\citeauthoryear{Hosokawa \& Inutsuka}{2005}]{Hosokawa:2005}
Hosokawa T., Inutsuka S., 2005, ApJ, 623, 917
\bibitem[\protect\citeauthoryear{Hosokawa \& Inutsuka}{2006}]{Hosokawa:2006}
Hosokawa T., Inutsuka S., 2006, ApJ, 646, 240
\bibitem[\protect\citeauthoryear{Kahn}{1954}]{Kahn:1954}
Kahn F.~D., 1954, Bull. Astron. Inst. Neth., 12, 187
\bibitem[\protect\citeauthoryear{Katz}{1992}]{Katz:1992}
Katz N., 1992, ApJ, 391, 502  
\bibitem[\protect\citeauthoryear{Kim, Kim \& Ostriker}{Kim et al.}{2018}]{Kim:2018}
Kim J.-G., Kim W.-T., Ostriker E.~C., 2018, ApJ, 859, 68  
\bibitem[\protect\citeauthoryear{Koyama \& Inutsuka}{2002}]{Koyama:2002}
Koyama H., Inutsuka S., 2002, ApJ, 564, L97
\bibitem[\protect\citeauthoryear{Krumholz, Stone \& Gardiner}{Krumholz et al.}{2007}]{Krumholz:2007}
Krumholz M.~R., Stone J.~M., Gardiner T.~A., 2007, ApJ, 671, 518  
\bibitem[\protect\citeauthoryear{Kupilas et al.}{2021}]{Kupilas:2021}
Kupilas M.~M.., Wareing C.~J., Pittard J.~M., Falle S.~A.~E.~G., 2021, MNRAS, 501, 3137
\bibitem[\protect\citeauthoryear{Mackey}{2012}]{Mackey:2012}
Mackey J., 2012, A\&A, 539, A147
\bibitem[\protect\citeauthoryear{Mackey et al.}{2015}]{Mackey:2015}
Mackey J., Gvaramadze V.~V., Mohamed S., Langer N., 2015, A\&A, 573, A10
\bibitem[\protect\citeauthoryear{Mellema et al.}{2006}]{Mellema:2006}
Mellema G., Iliev I.~T., Alvarez M.~A., Shapiro P.~R., 2006, New Ast., 11, 374
\bibitem[\protect\citeauthoryear{Mewe, Kaastra \& Liedahl}{Mewe et al.}{1995}]{Mewe:1995}
Mewe R., Kaastra J.~S., Liedahl D.~A., 1995, Legacy, 6, 16  
\bibitem[\protect\citeauthoryear{Osterbrock}{1989}]{Osterbrock:1989}
Osterbrock D.~E., 1989, Astrophysics of Gaseous Nebulae and Active
Galactic Nuclei (Mill Valley: University Science Books)  
\bibitem[\protect\citeauthoryear{Peters et al.}{2017}]{Peters:2017}
Peters T., Naab T., Walch S., Glover S.~C.~O., Girichidis P.,
Pellegrini E., Klessen R.~S., Wünsch R., Gatto A., Baczynski C., 2017,
MNRAS, 466, 3293
\bibitem[\protect\citeauthoryear{Pittard, Wareing \& Kupilas}{Pittard et al.}{2021}]{Pittard:2021} 
Pittard J.~M., Wareing C.~J., Kupilas M.~M., 2021, MNRAS, 508, 1768
\bibitem[\protect\citeauthoryear{Raga, Cantó \& Rodríguez}{Raga et al.}{2012}]{Raga:2012} 
Raga A.~C., Cantó J., Rodríguez L.~F., 2012, Rev. Mex. Astron. Astrofis., 48, 149
\bibitem[\protect\citeauthoryear{Rijkhorst et al.}{2006}]{Rijkhorst:2006}
Rijkhorst E.-J., Plewa T., Dubey A., Mellema G., 2006, A\&A, 452, 907  
\bibitem[\protect\citeauthoryear{Ritzerveld}{2005}]{Ritzerveld:2005}
Ritzerveld J., 2005, A\&A, 439, L23
\bibitem[\protect\citeauthoryear{Rogers \& Pittard}{2013}]{Rogers:2013}
Rogers H., Pittard J.~M., 2013, MNRAS, 431, 1337
\bibitem[\protect\citeauthoryear{Rosdahl et al.}{2015}]{Rosdahl:2015}
Rosdahl J., Schaye J., Teyssier R., Agertz O., 2015, MNRAS, 451, 34  
\bibitem[\protect\citeauthoryear{Shetty \& Ostriker}{2012}]{Shetty:2012}
Shetty R., Ostriker E.~C., 2012, ApJ, 754, 2
\bibitem[\protect\citeauthoryear{Spitzer}{1978}]{Spitzer:1978}
Spitzer L., 1978, Physical Processes in the Interstellar Medium,
Wiley-Interscience, New York  
\bibitem[\protect\citeauthoryear{Str\"{o}mgren}{1939}]{Stromgren:1939}
Str\"{o}mgren B., 1939, ApJ, 89, 526
\bibitem[\protect\citeauthoryear{Sutherland}{2010}]{Sutherland:2010}
Sutherland R.~S., 2010, Ap\&SS, 327, 173
\bibitem[\protect\citeauthoryear{Vazquez-Semadeni et al.}{2007}]{Vazquez-Semadeni:2007}
Vazquez-Semadeni E., Gómez G.~C., Jappsen A.~K., Ballesteros-Paredes
J., González R.~F., Klessen R.~S., 2007, ApJ, 657, 870
\bibitem[\protect\citeauthoryear{Wareing, Pittard \& Falle}{Wareing et al.}{2017a}]{Wareing:2017a}
Wareing C.~J., Pittard J.~M., Falle S.~A.~E.~G., 2017a, MNRAS, 465, 2757
\bibitem[\protect\citeauthoryear{Wareing, Pittard \& Falle}{Wareing et al.}{2017b}]{Wareing:2017b}
Wareing C.~J., Pittard J.~M., Falle S.~A.~E.~G., 2017b, MNRAS, 470, 2283
\bibitem[\protect\citeauthoryear{Wareing et al.}{2018}]{Wareing:2018}
Wareing C.~J., Pittard J.~M., Wright N.~J., Falle S.~A.~E.~G., 2018, MNRAS, 475, 3598
\bibitem[\protect\citeauthoryear{Whalen \& Norman}{2006}]{Whalen:2006}
Whalen D., Norman M.~L., 2006, ApJS, 162, 281  
\end{thebibliography}





\bsp	
\label{lastpage}
\end{document}